\documentclass[12pt]{article}
\pdfoutput =1
\usepackage{graphics}
\usepackage{graphicx}
\DeclareGraphicsExtensions{.pdf}
\usepackage{float} %Include figure filesusepackage{graphicx} %Include figure files
\usepackage{subfloat}
\usepackage{amsmath}
\usepackage{slashed}
\usepackage{amsfonts}  
\usepackage{amssymb}

\usepackage{color}

\usepackage{subcaption}
\begin{document}
\date{}
%%%%%%%%%%%%%%%%%%%%
\title{{\bf{\Large Topology of black hole phase transition in JT gravity}}}
%%%%%%%%%%%%%%%%%%%%
\author{
{\bf { Hemant Rathi}$
$\thanks{E-mail:  hemant.rathi@saha.ac.in}}\\
 {\normalsize  Saha Institute of Nuclear Physics, 1/AF Bidhannagar, }\\
  {\normalsize Kolkata 700064, India }
  \\[0.3cm]
 {\bf { ~Dibakar Roychowdhury}$
$\thanks{E-mail:  dibakar.roychowdhury@ph.iitr.ac.in}}\\
 {\normalsize  Department of Physics, Indian Institute of Technology Roorkee,}\\
  {\normalsize Roorkee 247667, Uttarakhand, India}
\\[0.3cm]
}
\maketitle
%%%%%%%%%%%%%%%%%%%%%%%%%%%%%%%%%%%%%%%%%%%%%%%%%%%%%%%%%%%%
\abstract{We present a JT gravity setup coupled with $U(1)$ and $SU(2)$ Yang-Mills fields in two dimensions that reveals the onset of a small black hole to large black hole phase transition at finite chemical potential(s). We identify these black hole solutions as ``topological defects'' in the thermodynamic phase space and calculate the associated topological numbers following the standard procedure. We confirm the robustness of our model by estimating perturbative corrections to the bulk free energy at an arbitrary order in the Yang-Mills coupling. We also construct the Schwarzian for the boundary theory using the 2D gravitational action in the bulk and comment on the dual SYK like model where similar observations can be made.}
%%%%%%%%%%%%%%%%%%%%%%%%%%%%%%%%%%%%
\section{Introduction and motivation }

Understanding the thermodynamics of black holes, notably the Hawking-Page (HP) transition \cite{Hawking:1982dh} in the context of gauge/gravity duality \cite{Maldacena:1997re}-\cite{Gubser:1998bc}, is one of the crucial problems in theoretical physics. The story has its root in 1983 when the authors in \cite{Hawking:1982dh} had explored the thermodynamics of black holes in Anti-de Sitter (AdS) spacetime and observed a phase transition between the thermal radiation and black holes of different masses known as the HP transition. Following their discovery, it was further argued that the HP transition could be thought of as confinement/deconfinement transition \cite{Witten:1998zw} in the language of the dual $\mathcal{N}=4$ SYM living on the boundary.    

Recently, the authors in \cite{Wei:2021vdx}-\cite{Wei:2020rbh} propose an alternative view for the thermodynamics of black holes. In particular, they introduce the concept of topology analogous to the Duan's topological current $\phi$-mapping theory \cite{Duan:1979ucg}-\cite{Duantwo}. This approach involves identifying ``topological defects'' in the thermodynamical phase space accompanied by a computation of topological charges associated with these defects.

To be precise, the authors in \cite{Wei:2021vdx} investigate the topological properties of the charged AdS  black holes \cite{Chamblin:1999tk} (that allows a small to large black hole phase transition) and Born-Infeld AdS black holes \cite{Fernando:2003tz} in four spacetime dimensions. They define appropriate thermodynamic function using the temperature and the entropy of the system and compute the topological charges for both small and large black holes. Their observation reveals that these black holes possess different topological charges, indicating that they belong to different topological classes.   

On the other hand, the authors in \cite{Wei:2022dzw} implement the concept of ``generalized free energy'' in order to define the vector field ($\phi$). They study the Schwarzschild, Reissner-Nordstr\"{o}m and Reissner-Nordstr\"{o}m (RN)-AdS black holes in 4D. In particular, they identify these black hole solutions as topological defects in thermodynamic phase space and compute the associated winding numbers. Their analysis reveals positive (negative) winding number that corresponds to the thermodynamically stable (unstable) black holes. They further use these topological properties to confirm a small to large (RN-AdS) black hole phase transition in 4D. 

In a related approach, the authors in \cite{Yerra:2022coh} employ the Bragg-Williams \cite{BW1} method to construct the off-shell free energy and estimate the topological charges associated with HP transition in charged AdS black holes. In this approach, one constructs the off-shell free energy using an order parameter and investigates their saddle points. Moreover, they show that the topological charges associated with HP transition match with the topological charges for the confinement/deconfinement transition for the dual QFT living on the boundary.

Despite the plethora of examples as alluded to the above, the literature on topological aspects of 2D black holes in the context of Jackiw-Teitelboim (JT) gravity \cite{Jackiw:1984je}-\cite{Teitelboim:1983ux} is lacking\footnote{See \cite{Lemos:1996bq}-\cite{Lemos:1993py}, for the black holes in 2D gravity.}. The theory is a 2D version of Einstein's gravity coupled with dilaton in the presence of a negative cosmological constant. Notably, JT gravity supports the HP transition when coupled with $U(1)$ and $SU(2)$ Yang-Mills (YM) fields \cite{Lala:2020lge}. It turns out that the presence of non-trivial gauge interactions result in a richer phase structure of 2D black holes, that also includes signatures of HP transition \cite{Lala:2020lge}. Therefore, exploring the topological aspects of 2D black holes in the presence of non-trivial gauge interactions is an important direction to be pursued.   

It is noteworthy to mention that the JT gravity allows a dual description in terms of the Sachdev-Ye-Kitaev (SYK) model \cite{Sachdev:1992fk}-\cite{Gaikwad:2018dfc} whose ``massless'' sector matches with the Schwarzian computed in a JT gravity\footnote{For comprehensive reviews on JT/SYK duality, see \cite{Rathi:2024qsy}-\cite{Sarosi:2017ykf}.} setup \cite{Maldacena:2016hyu}. The model SYK is a 1D quantum mechanical system consisting N Majorana fermions interacting simultaneously. The key feature of this model is that one can solve the 2-point and 4-point correlation functions exactly in the Large N limit. Notably, the authors in \cite{Zhang:2018kik}-\cite{Garcia-Garcia:2023uwh} explore some topological properties of the SYK model. However, as we point out, the SYK dual to our JT model is different from what has been studied so far.

The aim of this article is to explore the topological properties of JT gravity coupled with $U(1)$ and $SU(2)$ Yang-Mills gauge fields \cite{Lala:2020lge}. To be specific, we adopt the concepts of generalized free energy and winding numbers \cite{Wei:2022dzw} and using them we study topological defects in thermodynamic phase space of 2D black holes. Our analysis provides concrete evidence for a phase transition between a small and a large black hole. It turns out that the small black hole carries a negative topological charge, which is therefore unstable. On the other hand, the winding number corresponding to the large black hole turns out to be positive which indicates a stable phase. Moreover, our model is robust in the sense that the phase transition persist at all order in the YM gauge couplings. Finally, we calculate the Schwarzian for the boundary theory which shows the path to construct the dual SYK model.

The organisation for rest of the paper is as follows.

$\bullet$ In Section 2, we briefly review the thermodynamics of 2D black holes in the presence of non-trivial gauge interactions \cite{Lala:2020lge}.

$\bullet$ In Section 3, we introduce the concept of generalized free energy and defects \cite{Wei:2022dzw} in the vector space and identify them with the small and large black holes. Finally, we compute the associated winding numbers that reveals a phase transition between a small and a large black hole. 

$\bullet$ In Section 4, we show the existence of the HP transition in 2D gravity at an arbitrary order in the YM gauge couplings.

$\bullet$ In Section 5, we construct the boundary Schwarzian and discuss the dual SYK model at finite chemical potentials.

$\bullet$ In Section 6, we finally conclude our discussion and point out some interesting future directions.

\section{Thermodynamics of 2D black holes}\label{thm2d}

In this Section, we review some of the essential features of HP transition \cite{Hawking:1982dh} in a JT gravity set up as presented by the authors in \cite{Lala:2020lge}. The 2D gravity action can be obtained following a dimensional reduction of five dimensional Einstein's gravity coupled with $U(1)$ gauge fields as well as $SU(2)$ YM fields \cite{Fan:2015aia}, 

\begin{align}\label{5Daction}
S_{5D}=&\int d^{5}x \sqrt{-g_{(5)}}\Bigg(\mathcal{R}^{(5)}-3\Lambda-\frac{\kappa}{4g_{s}^{2}}
F^{a}_{MN}F^{aMN}-\frac{\xi}{4}F_{MN}F^{MN}-\nonumber\\
&\frac{\sigma}{2g_{2}^{2}}
\frac{\epsilon^{MNPQR}}{\sqrt{-g_{(5)}}}F^{a}_{MN}F^{a}_{PQ}A_{R}\Bigg),
\end{align}
where $(M,N,..)=(1,2..,5)$ are the 5D indices, $a=1,2,3$ denotes the non-Abelian indices, $\mathcal{R}^{(5)}$ denotes the Ricci scalar in 5D and $\Lambda$ is the negative cosmological constant. Furthermore, here $\xi$ and $\kappa$ are the respective dimensionless coupling constants associated with the $U(1)$ gauge fields and $SU(2)$ YM fields. 

The 2D model can be obtained following the reduction ansatz \cite{Lala:2020lge}
\begin{align}\label{dimredansatz}
ds_{(5)}^{2}&=\;ds_{(2)}^{2}+\Phi^{\frac{2}{3}}(x^{\mu})dx_{i}^{2},\quad ds_{(2)}^{2}=g_{\mu\nu}dx^{\mu}dx^{\nu},\nonumber\\
A_{M}^{a}dx^{M}&
=A_{\mu}^{a}dx^{\mu}, \quad A_{M}dx^{M}=A_{\mu}dx^{\mu},
\end{align}
where $\mu,\nu = (0,1) $ are the 2D indices of the bulk spacetime, $i=(2,3,4)$ denotes the compact directions and $\Phi(x^{\mu})$ is the dilaton of the 2D theory.

Substituting (\ref{dimredansatz}) into (\ref{5Daction}) and integrating out the compact directions, one finds
\begin{align}\label{2Dgravity}
S_{2D}&=\int d^{2}x\sqrt{-g_{(2)}}\;\Bigg[\mathcal{R}^{(2)}\Phi -3 \Lambda \Phi
-\Phi\Bigg(\frac{\kappa}{4g_{s}^{2}}
F^{a}_{\mu\nu}F^{a\mu\nu}+\frac{\xi}{4}F_{\mu\nu}F^{\mu\nu}\Bigg) \Bigg]+S_{bdy},
\end{align}
where $\mathcal{R}^{(2)}$ denotes the Ricci scalar of the 2D theory and $S_{bdy}=S_{GHY}+S_{ct}$ is the corresponding boundary term. Here, $S_{GHY}\sim\int dt\sqrt{-\gamma}\Phi\mathcal{K}$ is the standard Gibbons-Hawking-York term\footnote{Here, $\gamma_{tt}$ is the induced metric and $\mathcal{K}$ is the trace of the extrinsic curvature.} and $S_{ct} $ denotes the counter term. It should be noted that the parent 5D action (\ref{5Daction}) contains a non-vanishing Chern-Simons term with a coupling constant $\sigma$ which vanishes identically during the dimensional reduction.

The authors \cite{Lala:2020lge} compute the vacuum and black hole solutions associated with the 2D gravity model (\ref{2Dgravity}), where they use the static light cone gauge\footnote{In the light cone gauge, metric ansatz takes the form $ds^2=e^{2\omega(z)}\left(-dt^2+dz^2\right),$ along with the gauge fields $A_{\mu}=(A_t(z),0)$ and $A_{\mu}=\chi(z) \tau^3dt+\eta(z)\tau^1dz$, where $\tau=\sigma/2i$ are the Pauli metrics \cite{Lala:2020lge}. } and adopt a perturbative approach while treating $\xi$ and $\kappa$ as expansion parameters. 

The effective free energy of the system turns out to be
\begin{align}\label{freeenergy}
\Delta \mathcal{F}=\frac{1}{18\Lambda(\mu-1)}\Big[(\mu-1)\left\{-7Q^{2}\Lambda
\xi+16\xi\left(-30+0.22\mu\right)\right\}+120\Lambda\kappa\Big].
\end{align}

On the other hand, the average energy of the black hole phase in a canonical ensemble turns out to be
\begin{align}\label{avgeng}
\langle\mathcal{E}\rangle=\frac{1}{450\Lambda(\mu-1)^{2}}\left( 3000\kappa\Lambda(3\mu-1)-\xi(\mu-1)^{2}\left(12000
+175Q^{2}\Lambda+88\mu\right)\right)
\end{align}
where $Q$ is the $U(1)$ gauge charge and $\sqrt{\mu}$ is the location of the black hole horizon, which is related to the Hawking temperature of the black hole as, $T_H=\sqrt{\mu}/\pi$.

The entropy of the system can be obtained using Wald's prescription \cite{Wald:1993nt} ,
\begin{align}\label{waldentropy}
\begin{split}
S_{W}=&4\pi\Phi\big|_{\rho=\sqrt{\mu}}\\
=&-\frac{1}{96\Lambda\sqrt{\mu}(|\mu-1|)}\Big[(|\mu-1|)\Big\{-33Q^{2}\Lambda
\xi+64\mu\left(1+0.89\xi\right)   \\
&-15Q^{2}\xi\left(\log\mu\right)\Big\}-240
\kappa\Big].
\end{split}
\end{align}

It is interesting to notice that all the thermodynamic variables like the effective free energy (\ref{freeenergy}), average energy (\ref{avgeng}) and the thermal entropy (\ref{waldentropy}) of the system suffer from a discontinuity at $\mu=1$, indicating the onset of first-order phase transition \cite{Lala:2020lge} due to the presence of non-trivial $SU(2)$ YM interactions that comes with a coupling $\kappa$. In other words, the diverging terms appear to be proportional to the YM coupling $\kappa$. Moreover, the analysis also reveals that below a critical temperature ($T<T_H$), the system is dominated by thermal radiation . However, on increasing the temperature ($T> T_H$) of the system, thermal radiation collapses to form an unstable black hole with positive free energy and negative heat capacity. On further increasing the temperature, the unstable black hole transits into a globally stable black hole phase with negative free energy and positive heat capacity \cite{Lala:2020lge}. 

\section{Topological phase transition in 2D gravity}
In this Section, we probe into the topological aspects of 2D black holes in a gauge interacting model of JT gravity (\ref{2Dgravity}). Our analysis follows closely that of \cite{Wei:2022dzw}, where we identify black holes as topological defects in thermodynamic phase space. This approach has its analogue in various novel phenomena pertaining to condensed matter systems, for example, the Quantum Hall effect \cite{Halleffect}.

To begin with, we introduce a vector field $\phi(\Vec{x})$ and compute all its zero points $\Vec{x}=\Vec{z}$ such that $\phi(\Vec{x})\Big|_{\Vec{x}=\Vec{z}}=0$.  The zero points at $\Vec{x}=\Vec{z}$ are called the ``topological defects'' in the vector space, which will be identified with black hole solutions of (\ref{2Dgravity}).

The vector field is defined as \cite{Wei:2022dzw}
\begin{align}
    \phi=\left(\frac{\partial \mathcal{F}_g}{\partial r},-\cot\theta\csc\theta\right),\hspace{2mm}0\leq\theta\leq\pi,
\end{align}
where $r$ is the location of the black hole horizon\footnote{In our calculation, we define $\sqrt{\mu}=r$. } and $\mathcal{F}_g$ stands for the ``generalised free energy'' of the system 
\begin{align}\label{genfg}
     \mathcal{F}_g=E-\frac{S}{\tau},
\end{align}
where $E$ and $S$ respectively denote the energy and entropy of the 2D black hole.

Here, $\tau$ is the free parameter having the dimension of the inverse temperature. It should be noted that $\mathcal{F}_g$ defined above (\ref{genfg}) is an ``off-shell'' free energy. However, one can make it ``on-shell'' following an identification with the inverse Hawking temperature of the black hole namely, $\tau=T_H^{-1}$. 

For the JT gravity model (\ref{2Dgravity}), one can readily write down the components of $\phi$ using the on-shell free energy (\ref{freeenergy}), which yields
\begin{align}
    \phi^r=&\frac{1}{\Lambda  \left(r^2-1\right)^2}\left(0.39 \xi  r^5-0.78 \xi  r^3-13.33 \kappa  \Lambda  r+0.39 \xi 
   r\right),\label{phir}\\
   \phi^{\theta}=&-\cot\theta\csc\theta.\label{phitheta}
\end{align}
Notice that, the component $\phi^{\theta}$ vanishes at $\theta=\pi/2$, meaning that $\pi/2$ is the zero point of $\phi^{\theta}$ . On the other hand, $\phi^{\theta}$ diverges at $\theta=\pi$, indicating that the vectors at $\pi$ will spread and point outwards in the phase space.  

The zero points associated with $\phi^r$ can be obtained by setting, $\phi^r=0$ which yields the following roots
\begin{align}\label{rzeros}
    r=0,\hspace{1mm}\pm0.3 \sqrt{11-\frac{64.22 \sqrt{\kappa } \sqrt{-\Lambda }}{\sqrt{\xi
   }}},\hspace{1mm}\pm\sqrt{\frac{5.83 \sqrt{\kappa } \sqrt{-\Lambda }}{\sqrt{\xi }}+1}.
\end{align}

\begin{figure}[htp]
\begin{center}
\includegraphics[scale=.45]{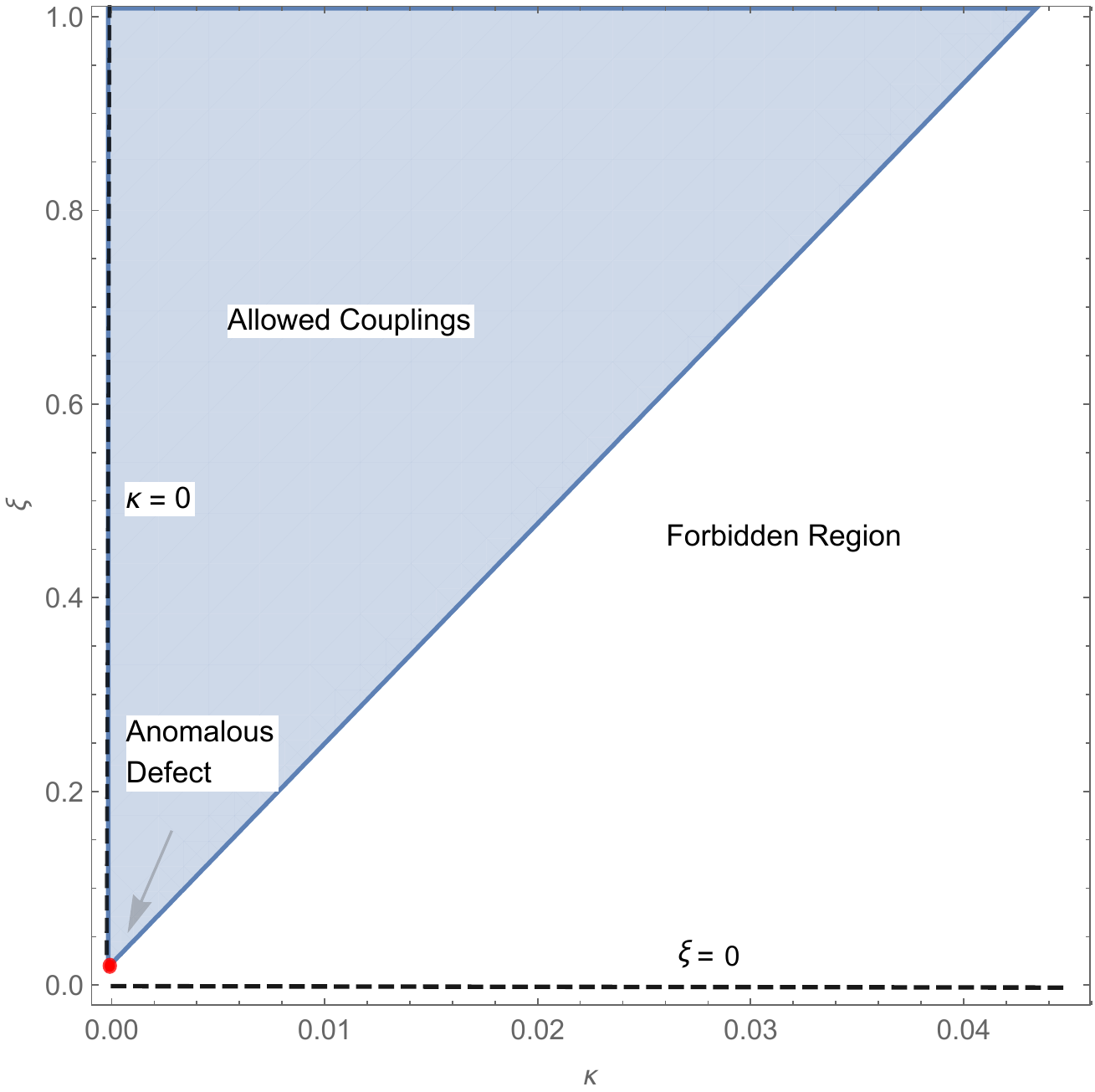}
\caption{$\xi$-$\kappa$ plot. Allowed values of the couplings are shown in the blue shaded region, and the white region is forbidden.} 
\label{couplings}
\end{center}
\end{figure}

Clearly, $r=0$ is the trivial solution. On the other hand, one must avoid possibilities of negative as well as imaginary roots that are being nonphysical. In other words, we consider finite and positive (real) roots. As we argue, an arbitrary choice of couplings ($\xi$ and $\kappa$) can result into nonphysical roots. A closer look further reveals that for roots \eqref{rzeros} to be real, the gauge couplings $\xi$ and $\kappa$ must satisfy the bound\footnote{ For $\frac{\kappa}{\xi}=0.044$, the solution sort of becomes trivial. We are left with two ``zero'' roots and one non-zero positive root $r=1.414$. Once the bound is crossed, one of the roots becomes imaginary.} $0<\frac{\kappa}{\xi}\leq0.044$ as shown in Figure \ref{couplings}. Therefore, one should treat the ratio $\frac{\kappa}{\xi}$ as an ``effective'' coupling of the theory, that is well compatible with the perturbative techniques adopted in this paper.

\begin{figure}[htp]
\begin{subfigure}{.5\textwidth}
  \centering
  % include first image
  \includegraphics[width=1\linewidth]{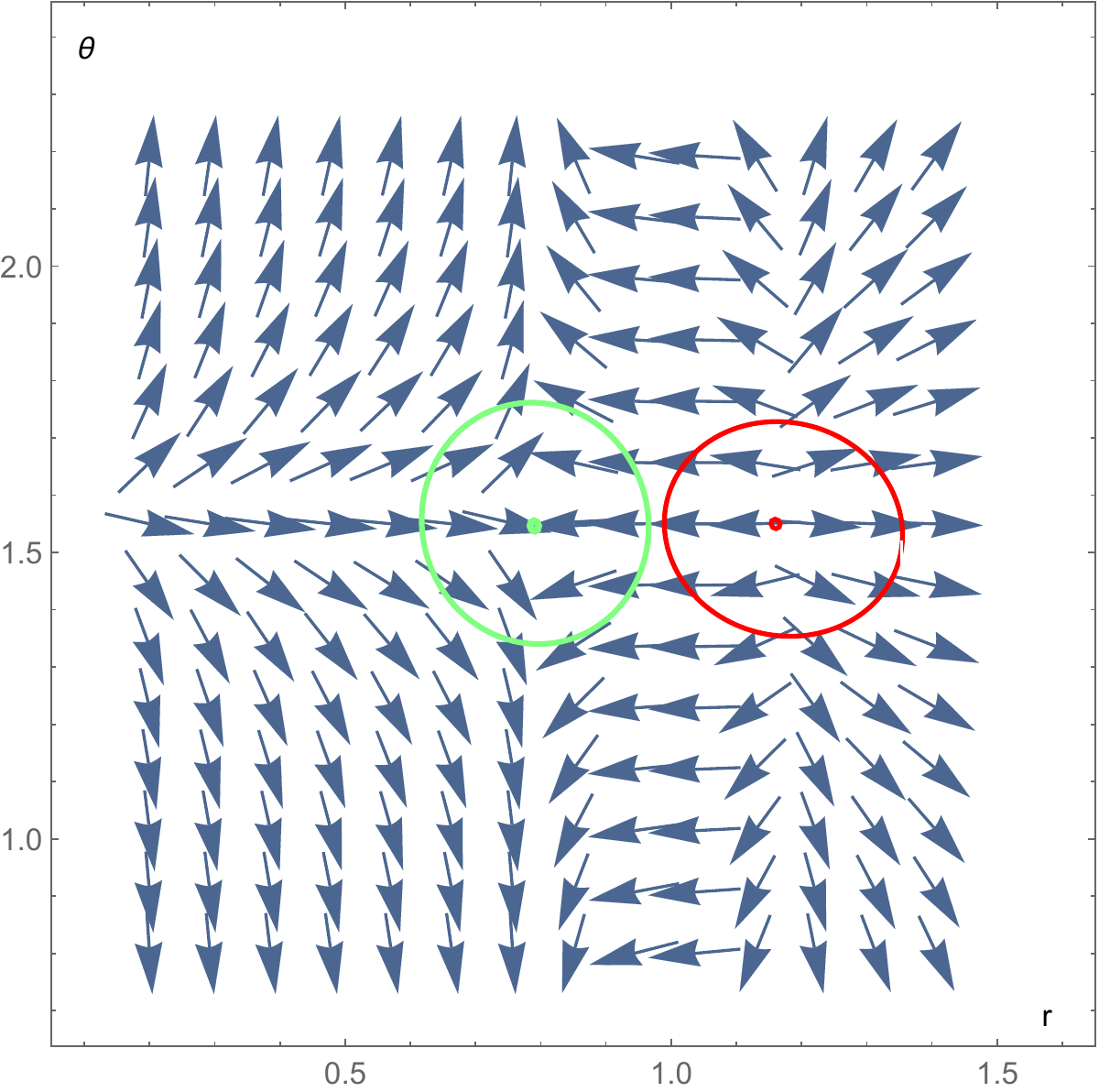} 
  \caption{$r-\theta$ plot for $\kappa=0.005$ and $\xi=0.8$ } 
  \label{imgvector}
\end{subfigure}
~~
\begin{subfigure}{.5\textwidth}
  \centering
  % include second image
  \includegraphics[width=1\linewidth]{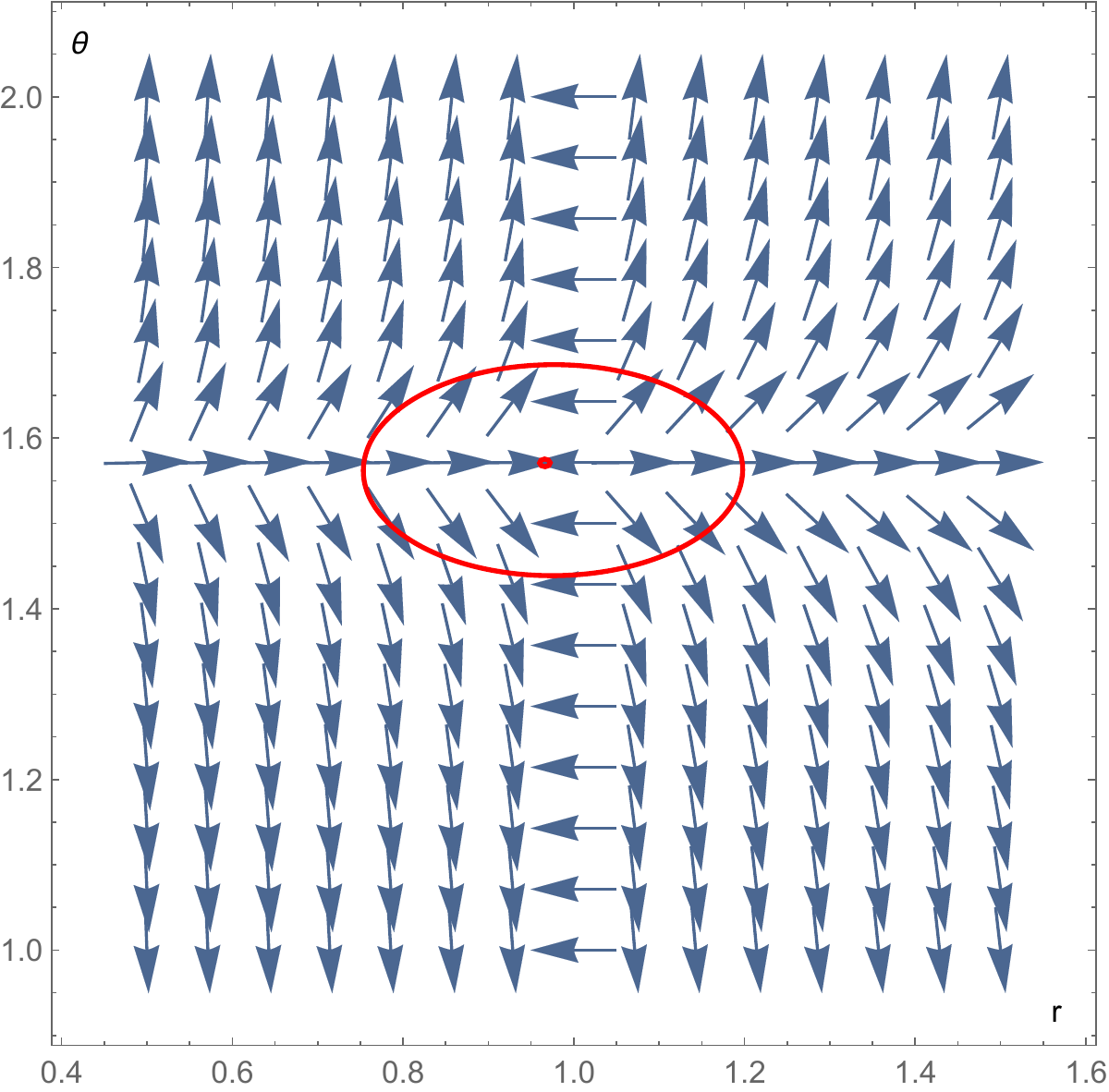}  
  \caption{$r-\theta$ plot for $\kappa=0$ and $\xi=0.8$ }
  \label{imgveczero}
\end{subfigure}
\caption{Vector plots for different values of $\xi$ and $\kappa$ }
\label{figurevectors}
\end{figure}

Next, we obtain vector plots for the 2D black hole system using $\phi^r$ (\ref{phir}) and $\phi^{\theta}$ (\ref{phitheta}) as shown in Figure \ref{figurevectors}. Figure \ref{imgvector} represents the vector plot for the couplings $\xi=0.8$ and $\kappa=0.005$ that satisfy the above bound.  Notice that, for this choice of couplings, equation (\ref{rzeros}) gives two physical roots namely, $r_1=0.78$ and $r_2=1.17$. In the vector plot \ref{imgvector}, we represent the first root by green colour and other by red colour. These defects represent a small and a large size black hole respectively.

In order to understand the thermodynamic stability associated with these defects, we compute the associated winding numbers ($w$). We find that the winding number for the smaller black hole is $-1$, while it is $+1$ for the larger black hole. The negative winding number indicates that the smaller black hole is thermodynamically unstable, while the positive winding number indicates that the larger black hole is thermodynamically stable \cite{Wei:2022dzw}. 

On the other hand, if we switch off the YM gauge coupling $\kappa$ for finite $\xi$ (which means setting the effective coupling, $\frac{\kappa}{\xi}=0$), we find that one of these defects goes away, and we are left only with one topological defect located at ($r=1, \theta=\pi/2$), as shown in Figure \ref{imgveczero}. At this critical point, the winding number diverges\footnote{In the limit $\kappa\rightarrow0$ and $\xi$ finite, one is left with only one (real) non-zero root $r=1$. It turns out that the Jacobi tensor $J^0=\frac{1}{\left(\text{r}^2-1\right)^3}\Big[0.469333 \left(\text{r}^6-3 \text{r}^4+3\text{r}^2-1\right) \csc \theta  \left(\cot ^2\theta +\csc ^2\theta \right)\Big]$ diverges at the critical point ($r=1, \theta=\pi/2$), which results in the diverging topological charge \cite{Wei:2022dzw}.} and therefore the topological interpretation of the defect becomes obscure. In other words, the critical point in the limit of the vanishing effective coupling ($\frac{\kappa}{\xi}\rightarrow0$), results in an ``anomalous defect'' in the thermodynamics phase space. To summarise, the topological properties of 2D black holes are highly sensitive to the choices of the gauge couplings $\xi$ and $\kappa$ and a physical interpretation of the topological defects can be given only in the presence of non-vanishing YM coupling ($\kappa\neq0$) at finite $\xi$.

\begin{figure}[htp]
\begin{subfigure}{.5\textwidth}
  \centering
  % include first image
  \includegraphics[width=1\linewidth]{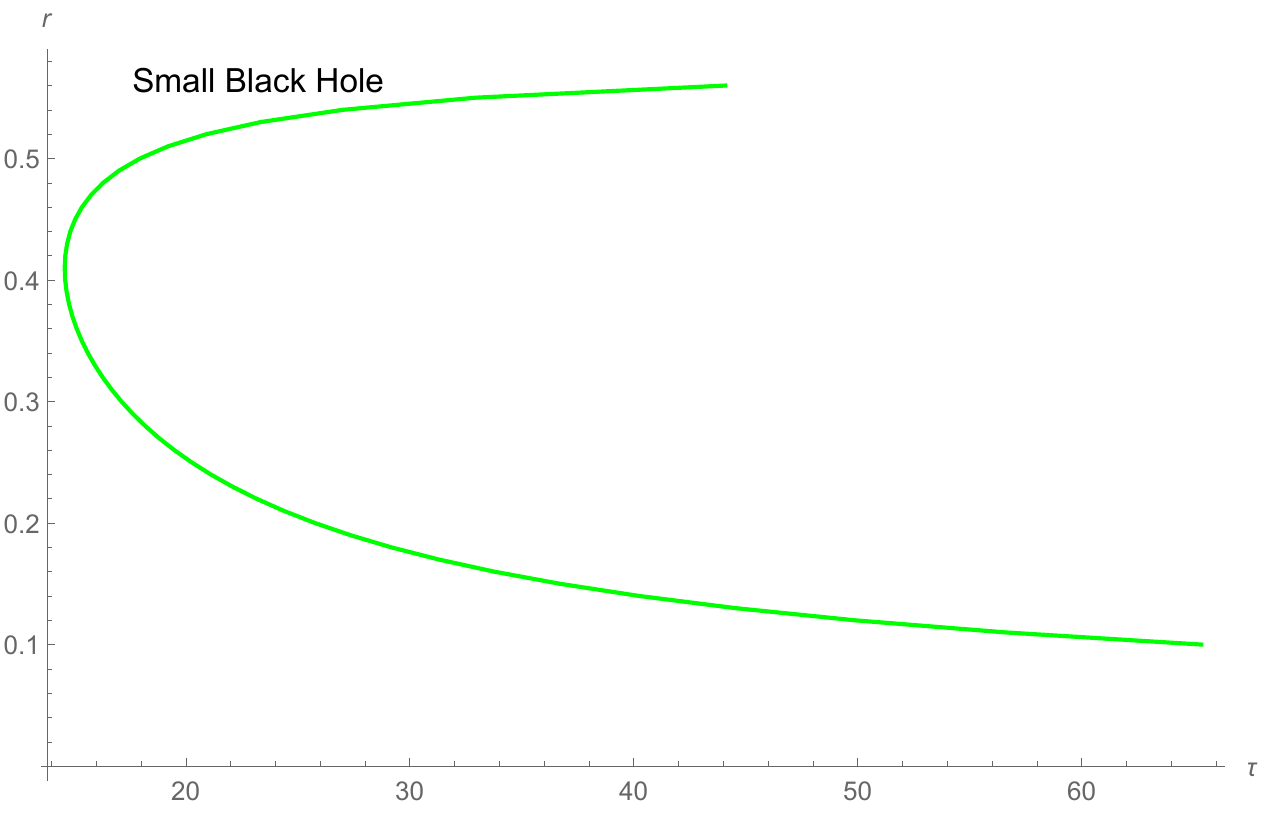} 
  %\caption{$r-\theta$ plot for $\kappa=0.005$ and $\xi=0.8$ } 
  \label{imglbh}
\end{subfigure}
~~
\begin{subfigure}{.5\textwidth}
  \centering
  % include second image
  \includegraphics[width=1\linewidth]{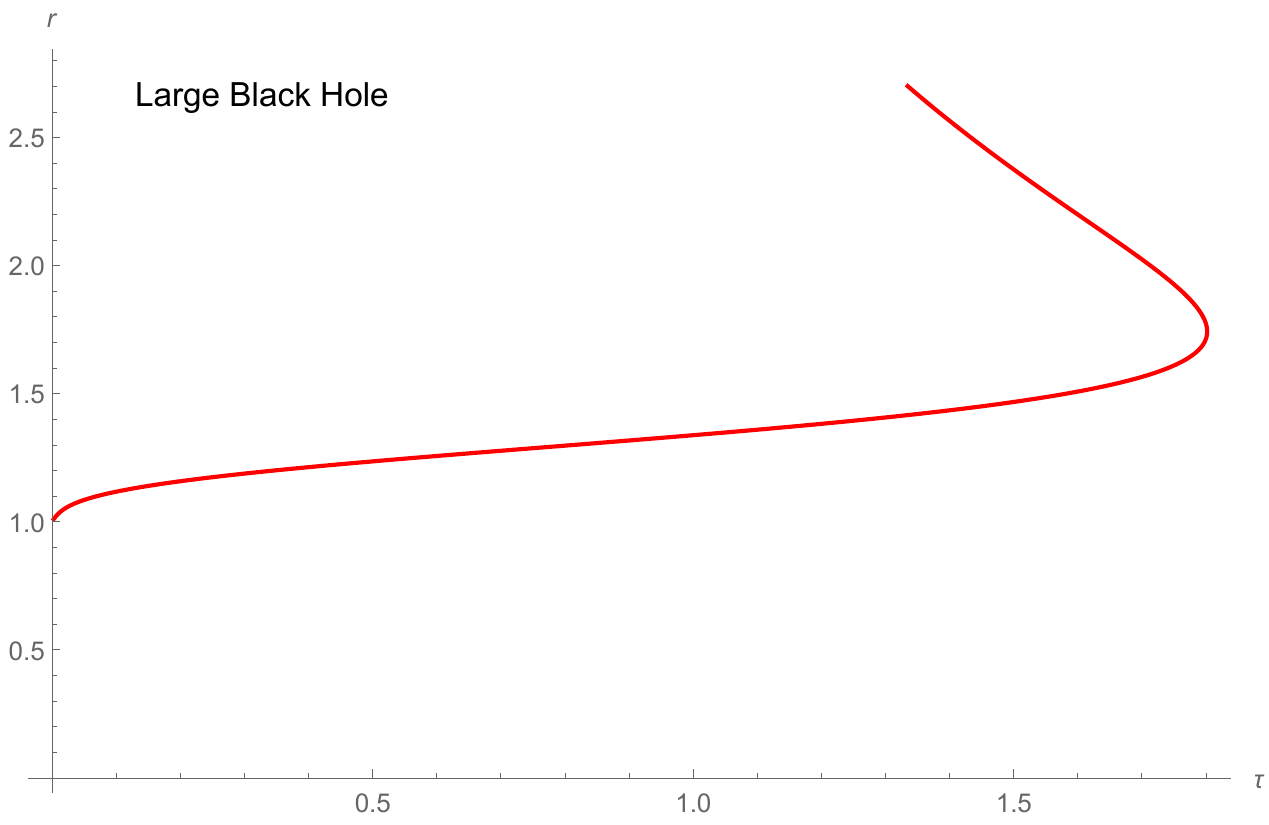}  
  %\caption{$r-\theta$ plot for $\kappa=0$ and $\xi=0.8$ }
  \label{imgsbh}
\end{subfigure}
\caption{$r-\tau$ plots for $\kappa=0.005$ and $\xi=0.8$ }
\label{figurebh}
\end{figure}

Aiming to uncover the possibilities for a small to a large black hole transition, we plot $r$ against the parameter $\tau$ setting $\kappa=0.005$ and $\xi=0.8$, as shown in Figure \ref{figurebh}. Our observation reveals that small black holes ($r<1$) appear at higher values of $\tau$ (or lower temperature), as shown in green, and these black holes are unstable, having negative winding number ($w=-1$). However, as one increases the temperature (lower $\tau$) of the system, the unstable black hole transits into a larger size ($r>1$) stable black hole, as shown in red, that corresponds to a positive winding number ($w=+1$). 

\section{Phase transition at higher order}
Clearly, the topological defects are highly sensitive to the choices of the gauge couplings ($\xi$ and $\kappa$) and in fact a non zero YM coupling ($\kappa\neq0$) is essential to trigger a small to large black hole phase transition, as alluded to the above. At this stage, one might wonder whether the above finding is an artifact of the perturbative expansion at LO in the gauge couplings ($\xi$ and $\kappa$). However, as we argue below, all the above features of the phase transition and hence the associated topological structure should persist at ``all'' order in the perturbation theory. 

To begin with, we expand all the background fields up to quadratic order in the gauge couplings $\xi$ and $\kappa$ as 
\begin{align} \label{quadexp}
    \mathcal{A}(z)=\mathcal{A}_0(z)+\xi\mathcal{A}_1^{ab}(z)+\kappa\mathcal{A}_1^{na}(z)+\xi^2\mathcal{A}_2^{ab}(z)+\kappa^2\mathcal{A}_2^{na}(z)+\xi \kappa\mathcal{A}_2^{abna}(z)+...,
\end{align}
where $\mathcal{A}(z)$ collectively  denotes the fields $\Phi$, $\omega$, $\chi$ and $\eta$ (see footnote 3 on page 4). Here, the subscript `$0$' denotes the pure JT gravity solutions, `$1$' and `$2$' stand for the leading and quadratic order corrections over pure JT gravity, and so on. Moreover, the superscripts `$ab$' and `$na$' denote contributions coming from the Abelian and non-Abelian gauge interactions. 

Using (\ref{2Dgravity}) and (\ref{quadexp}), one can immediately write down the general structure of the free energy at quadratic order

\begin{align}\label{genquadfe}
    \mathcal{F}^{(2)}=&\beta\int dz\Bigg[\xi^2\Bigg\{-2\Big(\omega_0''\Phi_2^{ab}+(\omega_1^{ab})''\Phi_1^{ab}+(\omega_2^{ab})''\Phi_0\Big)-3\Lambda e^{2\omega_0}\Big(\Phi_2^{ab}+2\Phi_1^{ab}\omega_1^{ab}\nonumber\\
    &+2\Phi_0(\omega_1^{ab})^2+2\Phi_0\omega_2^{ab}\Big)+\frac{Q^2}{2\Phi_0}e^{2\omega_0}\Bigg(2\omega_1^{ab}-\frac{\Phi_1^{ab}}{\Phi_0}\Bigg)\Bigg\}+\kappa^2\Bigg\{-2\Big(\omega_0''\Phi_2^{na}+\nonumber\\
    &(\omega_1^{na})''\Phi_1^{na}+(\omega_2^{na})''\Phi_0\Big)-3\Lambda e^{2\omega_0}\Big(\Phi_2^{na}+2\Phi_1^{na}\omega_1^{na}+2\Phi_0(\omega_1^{na})^2+2\Phi_0\omega_2^{na}\Big)+\nonumber\\
    &\frac{e^{-2\omega_0}}{2}\Big(2\eta_0\eta_1^{na}\Phi_0\chi_0^2+\eta_0^2\Phi_1^{na}\chi_0^2+2\eta_0^2\Phi_0\chi_0\chi_1^{na}-2\eta_0^2\Phi_0\chi_0^2\omega_1^{na}+\Phi_1^{na}\chi_0'^2-\nonumber\\
&2\Phi_0\omega_1^{na}\chi_0'{}^2+2\Phi_0\chi_0'\chi_1'^{na}\Big)\Bigg\}+\xi\kappa\Bigg\{-2\Big(\omega_0''\Phi_2^{abna}+(\omega_1^{ab})''\Phi_1^{na}+(\omega_2^{abna})''\Phi_0+\nonumber\\
&(\omega_1^{na})''\Phi_1^{ab}\Big)-3\Lambda e^{2\omega_0}\Big(\Phi_2^{abna}+2\Phi_1^{na}\omega_1^{ab}+2\Phi_0\omega_2^{abna}+2\Phi_1^{ab}\omega_1^{na}+4\phi_0\omega_1^{ab}\omega_1^{na}\Big)\nonumber\\
&+\frac{Q^2}{2\Phi_0}e^{2\omega_0}\Bigg(2\omega_1^{na}-\frac{\Phi_1^{na}}{\Phi_0}\Bigg)+\frac{e^{-2\omega_0}}{2}\Big(2\eta_0\eta_1^{ab}\Phi_0\chi_0^2+\eta_0^2\Phi_1^{ab}\chi_0^2+2\eta_0^2\Phi_0\chi_0\chi_1^{ab}-\nonumber\\
&2\eta_0^2\Phi_0\chi_0^2\omega_1^{ab}+\Phi_1^{ab}\chi_0'^2-2\Phi_0\omega_1^{ab}\chi_0'{}^2+2\Phi_0\chi_0'\chi_1'^{ab}\Big)\Bigg\}\Bigg],
\end{align}
where $'$ denotes the derivative with respect to the variable $z$ and $\beta =T^{-1}_H$.

Notice that, the quadratic free energy (\ref{genquadfe}) contains terms similar to those appearing at LO in the YM coupling ($\kappa$)
\begin{align}\label{genlfe}
    \mathcal{F}^{(\kappa)}=&\kappa\beta\int dz\Bigg[-2\Big(\omega_0''\Phi_1^{na}+(\omega_1^{na})''\Phi_0\Big)-3\Lambda e^{2\omega_0}\Big(\Phi_1^{na}+2\Phi_0\omega_1^{na}\Big)+\frac{e^{-2\omega_0}}{2}\Big(\Phi_0\chi_0'^2\nonumber\\
    &+\Phi_0\eta_0^2\chi_0^2\Big)\Bigg].
\end{align}

This is the piece in the free energy that yields a pole\footnote{This is precisely the $r=1$ root which corresponds to the anomalous defect in the limit of the vanishing ``effective'' coupling ($\frac{\kappa}{\xi} \rightarrow 0$). In the presence of non-zero effective coupling ($\frac{\kappa}{\xi} \neq 0$), topological defects lie on either side of this $\mu =1$ pole (see discussion on page 7, below figure \ref{figurebh}). As we switch-off the effective coupling, the defects on either side merge at $\mu$ (or $r$)$=1$. Notice that, in this limit, there is no singularity in the free energy and therefore there is no notion of phase transition.} at $\mu =1$ (\ref{freeenergy}) and is responsible for a HP like transition in 2D gravity. Therefore, one can expect similar behaviour at NLO. In other words, the divergence in the free energy persists at quadratic order in the YM couplings ($\kappa$), and we expect a phase transition similar to that at LO.  The only non-trivial modifications would take place at the level of the roots (\ref{rzeros}) which will be shifted as a result of higher order gauge interactions.

One can write down a general expression of free energy at $n^{th}$-order in the YM coupling ($\kappa$) and find a structure similar to (\ref{genlfe}) with increasing number of terms
\begin{align}\label{gennfe}
    \mathcal{F}^{(\kappa^{n})}=&\kappa^{n}\beta\int dz\Bigg[-2\Big(\omega_0''\Phi_n^{na}+(\omega_n^{na})''\Phi_0\Big)-3\Lambda e^{2\omega_0}\Big(\Phi_n^{na}+2\Phi_0\omega_n^{na}\Big)+\nonumber\\
    &\frac{e^{-2\omega_0}}{2}\Big(\Phi_{n-1}\chi_0'^2+\Phi_{n-1}\eta_0^2\chi_0^2\Big)+...\Bigg].
\end{align}
Accumulating all the above facts together, it is quite suggestive to infer that the $\mu =1$ divergence persists at all orders in the perturbation theory, and the topological phase transition is a generic feature of 2D gravity coupled with YM fields.

\section{Comments on the dual SYK model}

The purpose of this Section is to construct the boundary theory that is dual to the JT gravity model (\ref{2Dgravity}). Notice that, the model (\ref{2Dgravity}) consists of a $U(1)$ and $SU(2)$ YM fields, which source two different chemical potentials for the boundary theory namely the chemical potential $\mu_{em}$ corresponding to electromagnetic $U(1)$ and the second chemical potential $\mu_{YM}$ is sourced due to the $U(1)$ subgroup of $SU(2)$. Since, gauge fields play crucial role in obtaining the topological structure in 2D gravity, therefore, it is natural to claim that these chemical potentials will contribute non-trivially to the topological properties of the dual SYK model. 

 Based on some notable work \cite{Sachdev:2015efa}-\cite{Gaikwad:2018dfc}, we propose the dual SYK model to be of the following form
\begin{align}
    S=\int d\tau\Bigg[\frac{1}{2}\psi_i^\dagger\left(\partial_\tau-\mu_{em}\right)\psi_i+\frac{1}{2}\psi_i^{(a)}{}^\dagger\left(\partial_\tau-\mu_{YM}\right)\psi_i^{(a)}-\mathcal{H}\Bigg],
    \end{align}
    where $\psi_i$ and $\psi_i^{(a)}$ are the complex fermions charged under different gauge groups. The superscript $a$ represents the $SU(2)$ indices and $\mathcal{H}$ is the interaction Hamiltonian.

In order to have further insights into the structure of $\mathcal{H}$, one has to construct the effective boundary action for the 2D gravity model (\ref{2Dgravity}), which can be obtained by substituting the equation of motion for the dilton ($\Phi$) back into the action \cite{Maldacena:2016upp}. Notice that, the action (\ref{2Dgravity}) is linear in $\Phi$, therefore the bulk part vanishes identically and the entire contribution comes due to the Gibbons-Hawking-York term \eqref{2Dgravity}. 

In order to extract the boundary contribution, we consider the asymptotic ($z\sim \epsilon\rightarrow0$) expansion of the Euclidean space-time metric \cite{Lala:2020lge} 
    \begin{align}\label{lee1}
        ds^{2}\approx\left(\frac{C}{z^3}+\frac{1}{z^2}\right)\left(dt^2+dz^2\right),
    \end{align}
where $C=\alpha \xi+\beta\kappa$ and ($\alpha,\beta$) are constants. 

Now, we parameterize the spacetime coordinates $(t,z)$ using the variable $u$ and express the line element (\ref{lee1}) as
    \begin{align}\label{lee2}
        ds^{2}\approx\left(\frac{C}{z(u)^3}+\frac{1}{z(u)^2}\right)\left(t'^2+z'^2\right)du^2,
    \end{align}
    where $'$ denotes the derivative with respect to $u$. 
    
    Moreover, we impose the following boundary condition on the dilaton
    \begin{align}\label{bcphi}
        \Phi\Big|_{bdy}=\Phi_r(u)\left(\frac{1}{\epsilon}+\frac{C}{\epsilon^2}\right),
    \end{align}
    where $\Phi_r(u)$ is the regularised value of $\Phi$ at the boundary and $\epsilon$ is the UV cut-off.

    Using (\ref{lee2})-(\ref{bcphi}), one can elegantly express the boundary action as 
    \begin{align}
        S_{b}=-\int du\Phi_r(u)\Big[\text{Sch}\left(t(u),u\right)-(\mu_{em}+\mu_{YM})\mathcal{M}\left(t(u),u\right)\Big],
    \end{align}
    where $t(u)$ is the field variable, and $\text{Sch}\left(t(u),u\right)$ is the usual Schawrzian derivative \cite{Maldacena:2016upp}. Moreover, here we absorb the gauge couplings in the respective chemical potentials $\mu_{em}$ and $\mu_{YM}$. Notice that, the boundary value ($\Phi_r$) of the dilaton serves as the coupling of the dual SYK model and the function $\mathcal{M}$ can be expressed as 
    \begin{align}\label{mexp}
        \mathcal{M}\left(t(u),u\right)=\hspace{1mm}&\frac{1}{2 \epsilon ^3 t'(u)^3}\Bigg[ t'(u) \left(\epsilon ^2 t''(u)^2+t'(u)-\epsilon ^2 t^{'''}(u) \left(t'(u)-1\right)\right)-\epsilon ^2 t''(u)^2+\nonumber\\
        &2  \epsilon ^2  t'(u) \left(t''(u)^2-t^{'''}(u) t'(u)\right)\Bigg].
    \end{align}

 Given (\ref{mexp}), the next non-trivial task is to identify the interaction Hamiltonian ($\mathcal{H}$) for the dual SYK model. One can construct the interaction Hamiltonian ($\mathcal{H}$) following a reverse engineering technique. In this approach, one must figure out the 2-point and 4-point correlations for a general $q^{th}$ order interaction ($\mathcal{H}$) \cite{Maldacena:2016upp} and write down the effective action in the deep IR limit. Following that, one needs to map terms in (\ref{mexp}) to those in the effective action, by means of suitable coordinate transformations and symmetries of theory and fix interaction terms accordingly. We leave this interesting direction for future investigations.

\section{Conclusion and future directions}
To summarise, we show that for a topological phase transition to occur in a JT gravity setup, the gauge couplings must satisfy the bound $0<\frac{\kappa}{\xi}\leq 0.044$. In the limit when $\kappa\rightarrow 0$ and $\xi$ is finite, one encounters a situation with no phase transition that can be associated with an ``anomalous'' defect in the thermodynamic phase space, where the corresponding winding number becomes large. When the bound is satisfied, we have an analogue of the HP transition where we identify topological defects as small and large black holes and obtain their associated winding numbers. 

We show that small black hole corresponds to a negative winding number and is therefore thermodynamically unstable. On the other hand, the large black hole is associated with positive winding number and is thermodynamically stable. Our analysis suggests that an unstable small-size black hole transits into a stable large-size black hole beyond a particular critical value of the parameter $\tau \left(=T_H^{-1}\right)$, confirming the existence of phase transition in 2D gravity.

In addition to the above, we also argue that the HP transition persists at ``all orders'' in the YM gauge couplings, rather than just being specific to the linear order effect in a perturbation theory. Finally, we propose an effective action for the dual QFT at large N, that has chemical potentials corresponding to the $U(1)$ and $SU(2)$ gauge groups in the bulk theory. Starting from the bulk theory, we construct the Schwarzian for the boundary theory that give us enough hint for the dual SYK like model that one might be able to construct in the near future. Along with this, we also outline some other projects that might be worth exploring in the future.

$\bullet$ Related to the above discussion, it would be an exciting project to figure out the interaction Hamiltonian for the dual boundary theory and explore its topological properties. Particularly, unveiling a topological transition on the SYK side of the duality would be an exciting direction to look for.

$\bullet$ It would be interesting to study the phase stability of 2D black holes in the presence of quartic interactions \cite{Rathi:2021aaw} and/or ModMax interactions \cite{Rathi:2023vhw} from the topological approach and demystify the richer phase structure.

$\bullet$ Finally, it would be interesting to understand the black hole to wormhole phase transition in two dimensions \cite{Rathi:2021mla} from the above topological perspective.  

\section*{Acknowledgments}
The authors are indebted to the authorities of Indian Institute of Technology, Roorkee for their unconditional support towards researches in basic sciences. The authors would like to convey their sincere thanks to Arindam Lala for his collaboration during the early stages of this project. HR would like to thank Arnab Kundu and the authorities of Saha Institute of Nuclear Physics, Kolkata, for their support. HR would like to thank Sarthak Parikh for his support at the Indian Institute of Technology, Delhi. DR would like to acknowledge The Royal Society, UK for financial assistance. DR also acknowledges the Mathematical Research Impact Centric Support (MATRICS) grant (MTR/2023/000005) received from ANRF, India.

\end{document}